\documentclass[12pt]{article}

\usepackage{amssymb,amsmath}
\usepackage{graphicx}

\newcommand{\be}{\begin{equation}}
\newcommand{\ee}{\end{equation}}
\newcommand{\Dlt}{\Delta}
\newcommand{\dlt}{\delta}
\newcommand{\prt}{\partial}
\newcommand{\br}{{\bf r}}

\newcommand{\bp}{{\bf p}}
\newcommand{\bv}{{\bf v}}
\newcommand{\bbe}{{\bf e}}
\newcommand{\bs}{{\bf s}}
\newcommand{\bu}{{\bf u}}

\newcommand{\bP}{{\bf P}}
\newcommand{\bt}{\beta}
\newcommand{\vp}{\varphi}
\newcommand{\ep}{\varepsilon}

\newcommand{\ra}{\rightarrow}
\newcommand{\sgm}{\sigma}

\newcommand{\om}{\omega}
\newcommand{\Om}{\Omega}

\newcommand{\dgr}{\dagger}

\newcommand{\cH}{{\cal H}}
\newcommand{\cM}{{\cal M}}
\newcommand{\vom}{\vec\omega}
\newcommand{\lgl}{\langle}
\newcommand{\rgl}{\rangle}

\begin{document}

\begin{center}

{\Large{\bf
Turbulent superfluid as continuous vortex mixture} \\ [5mm]
V.I. Yukalov} \\ [3mm]

{\it
Bogolubov Laboratory of Theoretical Physics, \\
Joint Institute for Nuclear Research, Dubna 141980, Russia \\
and \\
National Institute of Optics and Photonics,  \\
University of S\~ao Paulo, S\~ao Carlos 13560-970, Brazil}

\end{center}

\vskip 3cm

\begin{abstract}

A statistical model is advanced for describing quantum
turbulence in a superfluid system with Bose-Einstein condensate. Such a
turbulent superfluid can be realized for trapped Bose atoms subject to
either an alternating trapping potential or to an alternating magnetic
field modulating the atomic scattering length by means of Feshbach
resonance. The turbulent system is represented as a continuous mixture
of states each of which is characterized by its own vorticity
corresponding to a particular vortex.
\end{abstract}

\vskip 2cm



\newpage
\section{Quantum turbulence}

Turbulence in classical fluids has been intensively studied for many
years [1-5]. It can also occur in superfluids, such as $^4$He, $^3$He,
and trapped atomic gases. A superfluid differs from a classical fluid
in three principal respects: it exhibits two-fluid behavior; the
superfluid component can flow without dissipation; and flow of the
superfluid component is subject to quantum restrictions. One of the
major such restrictions is that the vortices in superfluids are
quantized [6-11]. Emphasizing that the peculiarities of turbulence in
superfluids, as compared to that in classical fluids, are due to
quantum effects, turbulence in superfluids is called {\it quantum
turbulence} [12-14].

There are several ways of producing turbulence in superfluids, e.g., by
realizing counterflows between the normal and superfluid components, by
moving grids or by vibrating objects immersed in the superfluid [12-14].
The fully developed quantum turbulence is represented by a {\it random 
complex tangle of quantized vortices}. Their appearance can be due to 
different reasons, to mutual friction between the normal and superfluid 
components [12-16], accompanied by the Kelvin-Helmholtz instability 
[17,18], to thermal and quantum tunneling [12], and to other dynamic 
instabilities [19,20]. When the external disturbance, producing 
turbulence, ceases to act on the fluid, turbulence decays through 
the Richardson cascades [12-14], the emission of sound radiation 
produced by vibrating vortices [21,22], by vortex reconnecion 
[23,24], and by Kelvin-wave cascades [25,26]. The evolution of 
quantized vortices can be visualized by injecting into the fluid 
small admixture particles [27].

Bose-Einstein condensed gases of trapped atoms [11, 28-33] provide
additional possibilities for studying superfluids in a well controlled
way. However, by rotating trapped Bose condensates, one can create
quantized vortices, which tend to form crystalline structures, when the
rotation velocity increases [34-36]. Simple rotation does not produce
turbulence in trapped systems. In order to create a tangle of vortices,
typical of quantum turbulence, it is necessary to use some other techniques.
Actually, a vortex is a particular case of coherent topological modes,
among which there exist many other soliton-like excitations [37,38]. Hence,
the general question is how it would be possible to create coherent
topological modes for trapped atomic systems?

A method for generating various coherent topological modes in traps has
been advanced, based on the modulation of atomic cloud density [39-41].
The modulation can be done in two ways, either by oscillating trapping
potentials [39-43] or by alternating the atomic scattering length using
Feshbach resonance techniques [33,44,45]]. Both these ways yield similar
results, so that it is a matter of convenience which of them to employ.
The generation of the coherent modes can be achieved by rather weak
modulation amplitudes, but provided that the frequency of the alternating
trapping potential or of the alternating scattering length is in resonance
with the transition frequency corresponding to the desired coherent mode
[39-45]. By increasing the modulation amplitude, it is feasible to realize
the nonresonant mode generation [42]. In particular, the generation,
without the direct resonance condition, can be done, e.g., by means of
parametric conversion or harmonic generation [46,47]. Very strong forced
oscillations can even destroy superfluidity at all [48]. The alternating
modulation of trapping potentials, to some extent, can be modelled by
combined rotations around several axes [49].

For theoretical description of quantum turbulence, it is sufficient to
consider the nonlinear Schr\"{o}dinger equation (Gross-Pitaevskii equation)
[49,50], keeping in mind low temperatures and weak atomic interactions, when
almost the whole system is in the Bose-condensed state. The generation of
coherent topological modes and, hence, turbulence, can also be realized under
more general conditions of finite temperatures and strong atomic interactions
[51]. The presence of the normal component is not necessary for the appearance
of quantized vortices. The latter arise because of the occurrence of a
heteroclinic instability [38] in a nonequilibrium system, when the homogeneous
atomic density becomes dynamically unstable with respect to its transformation
to a nonuniform density comprising topological defects, like quantized vortices.
This is analogous to the development of classical turbulence that is accompanied
by a strongly nonuniform distribution of density [52-55]. In quantum turbulence,
the nucleation of vortices, under the dynamic instability, is caused by
developing unstable collective excitations, such, e.g., as quadrupole modes
[19,20]. The typical feature of quantum turbulence is the random spatial
distribution of quantized vortices [12-14], similar to the models of dynamical
glass [56].

Experimentally, the modulation of the trapping potential has been used for the
generation of quantized vortices [57] and vortex tangles [58]. The appearance of
vortex tangles implies that the system enters the fully turbulent regime [12-14].

Turbulence is a strongly nonequilibrium phenomenon representing an essentially
nonuniform matter. This is why it is difficult for theoretical description. But,
in the case of fully developed stationary turbulence, it can be possible to
invoke statistical methods. It is the aim of the present paper to suggest a
statistical model of stationary quantum turbulence.

\section{Vortex mixture}

Quantized vortices are the necessary ingredients of quantum turbulence. Each
vortex can be represented as a filament passing through the fluid. The velocity
$\bf{v} = \bf{v}(\bf{r})$ at a point $\bf{r}$, due to a filament, is given by
the Biot-Savart formula
\be
\label{1}
\bv = \frac{n}{2m} \; \int \;
\frac{(\bs-\br)}{|\bs -\br|^3}\; \times d\bs \; ,
\ee
where $m$ is atomic mass, $n$ is an integer characterizing the quantum of
circulation, the vector $\bf{s}$ refers to a point on the filament, and the
integration is taken along the filament. The filament has a definite direction
associated with its vorticity
\be
\label{2}
\vec{\omega} \equiv \nabla \times \bf{v} \; .
\ee
Integrating velocity along a path around the vortex core gives the quantum of
circulation,
\be
\label{3}
\oint \bv \cdot d{\bf l} = \frac{2\pi n}{m}
\qquad (n=0,\pm 1,\pm 2,\ldots ) \; .
\ee
For example, in the case of a rectilinear vortex, with the vorticity along the
axis $z$, the velocity is
$$
\bv = \frac{n}{mr_\perp} \; \bbe_\vp = \frac{n}{mr^2_\perp}\; \bbe_z
\times \br_\perp \; ,
$$
where $r_{\perp} \equiv \sqrt{x^2 + y^2} = |\;\br_\perp|$. This gives the vorticity
$$
\vec\om_n = \frac{2\pi n}{m} \; \dlt(x) \dlt(y) \bbe_z \; .
$$
The case of $n=0$ means the absence of a vortex.

In the turbulent superfluid there are many vortices with different vorticities,
whose distribution can be characterized by a measure $m(\vec{\omega})$. Thus,
in the case of rectilinear vortices, integration of a function $f(\vec{\omega})$,
with a measure $m(\vec{\omega})$, gives
$$
\int f(\vec{\om}) \; dm(\vec{\om}) =
\sum_n \int  f(\vec{\om}_n) \; d\Om(\bbe_z) \; ,
$$
where summation is over the circulation quanta $n=0,1,2,\ldots$ and integration
is over the spherical angle around $e_z$, that is, over all directions of
vorticity.

A fluid with a quantized vortex represents a dynamic state that is principally
different from the state without vortices. Moreover, the states with the
vortices, having different winding numbers or vorticities can be treated as
different states. At any instant of time, the system, formed by multiple
regions, each containing a vortex, can be considered as a composition of states
corresponding to these different regions. So, at each instant of time, the
system is represented as a configuration, composed of regions with different
vortices. The density of vortices in the system and their directions are random
in space and are varying in time. This means that the vortex configuration
varies. The statistical description of such a situation assumes an averaged
picture corresponding to the averaging over all admissible state configurations.
The theory of describing such heterophase mixtures has been advanced in [59-63]
and reviewed in [64-66], where all mathematical details of the related averaging
procedure are elucidated.

Following this approach [64-66], we come to an effective Hamiltonian
characterizing the random mixture. In the present case, this Hamiltonian is
the direct integral
\be
\label{4}
\widetilde H = \int^\oplus H(\vom)\; dm(\vom) \; \bigoplus \; H_2  \; .
\ee
Here $H(\vec{\omega})$ is the Hamiltonian of a system with a vortex characterized
by the vorticity $\vec{\omega}$ and $H_2$ is the Hamiltonian corresponding to the
normal (nonsuperfluid) state. The latter has to be taken into account since,
in the process of generating turbulence, superfluidity can be destroyed in a
part of the system. This Hamiltonian (4) is defined on the {\it mixture space}
\be
\label{5}
\cM = \bigotimes_{\vec{\om}} \cH(\vec{\om}) \bigotimes \cH_2 \; ,
\ee
which is a continuous tensor product of the weighted Hilbert spaces. More
details on the meaning of continuous products are given in the Appendix. Each
weighted space ${\cal H}(\vec{\omega})$ is a copy of the system Hilbert space,
with a weighted scalar product [64-66], corresponding to the state with a vortex
labelled by its vorticity $\vec{\omega}$. And ${\cal H}_2$ is a space
corresponding to the normal (nonsuperfluid) state.

The superfluid states with the vortices are described by the grand
Hamiltonians
\be
\label{6}
 H(\vec{\om}) = \hat H(\vec{\om}) -
\mu_0 (\vec{\om}) N_0 (\vec{\om}) -
\mu_1 (\vec{\om}) \hat N_1 (\vec{\om}) \;  ,
\ee
in which $\hat{H}(\vec{\omega})$ is the energy operator, $N_0(\vec{\omega})$
is the number of condensed atoms, $\hat{N}_1(\vec{\omega})$ is the
number-of-particle operator for uncondensed atoms, and $\mu_0(\vec{\omega})$
and $\mu_1(\vec{\omega})$ are the Lagrange multipliers guaranteeing the
normalization conditions for the number of Bose-condensed atoms
$N_0(\vec{\omega})$ and for the number of uncondensed atoms
$N_1(\vec{\omega})$ in the superfluid state of vorticity $\vec{\omega}$
[51,67-69]. The nonsuperfluid state is characterized by the grand Hamiltonian
\be
\label{7}
 H_2 = \hat H_2 - \mu_2 \hat N_2 \; ,
\ee
where $\hat{H}_2$ is the corresponding energy operator, $\hat{N}_2$ is the
related number-of-particle operator, and $\mu_2$ is the chemical potential
of the normal state.

In what follows, we shall consider local atomic interactions
\be
\label{8}
 \Phi(\br) = \Phi_0 \dlt(\br) \; , \qquad \Phi_0 \equiv
4\pi \; \frac{a_s}{m} \; ,
\ee
where $m$ is atomic mass and $a_s$, scattering length. The local form (8)
is not principal, but is taken just for the brevity of notations. Throughout
the paper, the units are employed, where $\hbar = 1$ and $k_B = 1$.

The Hamiltonian energy operator in (6) has the form
$$
\hat H(\vec{\om}) = w_1(\vec{\om}) \int \hat \psi^\dgr (\vec{\om},\br)
\left [ \frac{\hat\bp^2}{2m} +
U(\br) \right ] \hat\psi(\vec{\om},\br) \; d\br \; +
$$
\be
\label{9}
 + \; \frac{w_1^2(\vec{\om})}{2} \;
\Phi_0 \int \hat\psi^\dgr(\vec{\om},\br)
\hat\psi^\dgr(\vec{\om},\br)
\hat\psi(\vec{\om},\br) \hat\psi(\vec{\om},\br) \; d\br \; ,
\ee
where the field operators $\hat{\psi}(\vec{\omega},\bf{r})$ act on the space
$\cal{H}(\vec{\omega})$ and the operator of momentum is $\hat{\bp}\equiv-i\nabla$.
The factor $w_1(\vec{\omega})$ is the geometric weight of the superfluid vortex
state with vorticity $\vec{\omega}$. The function $U(\bf r)$ is an external
potential, say, a trapping potential. The superfluid system is assumed to be
Bose-condensed. And the Bose-Einstein condensation necessarily requires that the
global gauge symmetry be broken [70].

The most convenient way of gauge symmetry breaking is by means of the
Bogolubov shift [71], which can be represented as the sum
\be
\label{10}
\hat\psi(\vec{\om},\br) = \eta(\vec{\om},\br)
\exp\{ i S(\br) \} + \psi_1( \vec{\om},\br) \; ,
\ee
in which $S(\bf r)$ is a real function, such that
\be
\label{11}
\bv = \bv(\br) = \frac{1}{m} \; \nabla S(\br)
\ee
is the superfluid velocity. The first term in (10) is the condensate wave
function and the second term is an operator of uncondensed atoms in the
superfluid state with vorticity $\vec{\omega}$. To exclude the double
counting, the terms in sum (10) are to be orthogonal to each other,
\be
\label{12}
\int \eta(\vec{\om},\br) \exp\{ iS(\br) \}
\hat\psi^\dgr_1(\vec{\om},\br)\; d\br = 0  \; .
\ee
And the statistical average for the operator of uncondensed atoms satisfies
the property
\be
\label{13}
\lgl \psi_1(\vec{\om},\br) \rgl = 0 \; ,
\ee
due to which the condensate wave function plays the role of the order parameter.

The Hamiltonian energy operator in (7) is
$$
\hat H_2 = w_2 \int \psi^\dgr_2(\br) \left [
\frac{\hat\bp^2}{2m} + U(\br) \right ] \psi_2(\br) \; d \br \; +
$$
\be
\label{14}
 + \; \frac{w_2^2}{2} \; \Phi_0 \int \psi_2^\dgr(\br)
\psi_2^\dgr(\br) \psi_2(\br) \psi_2(\br) \; d\br \;  ,
\ee
where the field operators $\psi_2(\bf{r})$, acting on ${\cal H}_2$, correspond
to the normal (nonsuperfluid) state and $w_2$ is the geometric weight of the
normal state. All field operators satisfy the Bose commutation relations.

The condensate wave function defines the density of condensed atoms in the
state with vorticity $\vec{\omega}$,
\be
\label{15}
\rho_0(\vec{\om},\br) = | \eta (\vec{\om},\br) |^2 \; .
\ee
The density of uncondensed atoms in the superfluid state, labelled by
$\vec{\omega}$, is
\be
\label{16}
 \rho_1(\vec{\om},\br) =  \lgl \psi_1^\dgr (\vec{\om},\br)
 \psi_1(\vec{\om},\br) \rgl \; .
\ee
And the density of the normal phase is
\be
\label{17}
 \rho_2(\br) =  \lgl \psi_2^\dgr(\br) \psi_2(\br) \rgl \; .
\ee

The number of condensed atoms in the $\vec{\omega}$ vortex state is
\be
\label{18}
N_0 (\vec{\om}) = w_1(\vec{\om}) \int \rho_0(\vec{\om},\br) \;
d\br \;  .
\ee
The number-of-particle operator of uncondensed atoms in that state is
\be
\label{19}
\hat N_1 (\vec{\om}) = w_1(\vec{\om}) \int
\psi_1^\dgr (\vec{\om},\br) \psi_1(\vec{\om},\br)\; d\br \; .
\ee
And the number-of-particle operator for atoms in the normal state is
\be
\label{20}
\hat N_2 = w_2 \int \psi_2^\dgr(\br) \psi_2(\br) \; d\br  \;  .
\ee
So that the number of uncondensed atoms in the $\vec{\omega}$ vortex
state reads as
\be
\label{21}
N_1 (\vec{\om}) \equiv \lgl \hat N_1(\vec{\om}) \rgl =
w_1(\vec{\om}) \int \rho_1(\vec{\om},\br) \; d\br \; ,
\ee
and the number of normal atoms being
\be
\label{22}
 N_2 \equiv \lgl \hat N_2 \rgl = w_2 \int \rho_2(\br) \; d\br \; .
\ee
The number of condensed atoms in all vortex states is the integral
\be
\label{23}
 N_0 = \int N_0(\vec{\om}) \; dm(\vec{\om}) \; ,
\ee
while the number of uncondensed atoms in all vortex states is
\be
\label{24}
N_1 = \int  N_1 (\vec{\om}) \; dm(\vec{\om}) \; .
\ee
The total number of atoms in the system is
\be
\label{25}
 N = N_0 + N_1 + N_2 \;  .
\ee

The geometric weight of the superfluid state with all types of vortices
reads as
\be
\label{26}
w_1 \equiv \int w_1(\vec{\om}) \; dm(\vec{\om}) \;  .
\ee
There can be in the system only two types of states, superfluid and normal,
which implies the normalization
\be
\label{27}
 w_1 + w_2 = 1 \qquad ( 0 \leq w_\nu \leq 1) \;  ,
\ee
where $\nu = 1,2$.

The equations of motion for the field variables are given by the related
variations: for the condensate wave function of a superfluid state with the
vorticity $\vec{\omega}$,
\be
\label{28}
i\; \frac{\prt}{\prt t} \; \eta(\vec{\om},\br,t) =
\lgl \; \frac{\dlt\widetilde H}{\dlt\eta^*(\vec{\om},\br,t)} \;
\rgl \; ;
\ee
for the field operator of uncondensed atoms in the superfluid state with the
vorticity $\vec{\omega}$,
\be
\label{29}
i\; \frac{\prt}{\prt t} \; \psi_1(\vec{\om},\br,t) =
\frac{\dlt\widetilde H}{\dlt\psi_1^\dgr(\vec{\om},\br,t)} \; ;
\ee
and for the field operator of atoms in the normal (nonsuperfluid) state,
\be
\label{30}
i\; \frac{\prt}{\prt t} \; \psi_2(\br,t) =
\frac{\dlt\widetilde H}{\dlt\psi_2^\dgr(\br,t)} \; .
\ee

Let us introduce the notations for the anomalous average
\be
\label{31}
\sgm_1(\vom,\br) \equiv
\lgl \psi_1(\vom,\br) \psi_1(\vom,\br) \rgl \;  ,
\ee
the anomalous triple correlator
\be
\label{32}
\xi(\vom,\br) \equiv
\lgl \psi_1^\dgr(\vom,\br) \psi_1(\vom,\br)
\psi_1(\vom,\br) \rgl \;   ,
\ee
and for the operator
$$
\hat X(\vom,\br) \equiv 2 \psi_1^\dgr(\vom,\br)
\psi_1(\vom,\br) \eta(\vom,\br) +
$$
\be
\label{33}
+ \eta^*(\vom,\br) \psi_1(\vom,\br) \psi_1(\vom,\br) +
\psi_1^\dgr(\vom,\br) \psi_1(\vom,\br) \psi_1(\vom,\br) \; ,
\ee
where, for brevity, the time dependence is not explicitly shown.

Then Eq. (28) yields the equation for the condensate wave function
$$
i\; \frac{\prt}{\prt t} \; \eta(\vom,\br) = w_1(\vom)
\left [ \frac{(\hat\bp + m\bv)^2}{2m} + U(\br) - \mu_0(\vom) \right ]
\eta(\vom,\br) \; +
$$
\be
\label{34}
+ \; w_1^2(\vom) \Phi_0 \left [ \rho_0(\vom,\br) \eta(\vom,\br)
+2\rho_1(\vom,\br) \eta(\vom,\br) +
\sgm_1(\vom,\br) \eta^*(\vom,\br) + \xi(\vom,\br)\right ] \;  .
\ee
Equation (29) results in the equation for the field operator of
uncondensed atoms
$$
i\; \frac{\prt}{\prt t} \; \psi_1(\vom,\br) = w_1(\vom)
\left [ \frac{\hat\bp^2}{2m} + U(\br) - \mu_1(\vom) \right ]
\psi_1(\vom,\br) \; +
$$
\be
\label{35}
+ \; w_1^2(\vom) \Phi_0 \left [ 2\rho_0(\vom,\br) \psi_1(\vom,\br)
+ \eta^2(\vom,\br) \psi_1^\dgr(\vom,\br) +  \hat X(\vom,\br)
\right ] \;   .
\ee
And Eq. (30) gives the equation for the field operator of normal atoms
\be
\label{36}
i\; \frac{\prt}{\prt t} \; \psi_2(\br) = w_2 \left [
\frac{\hat\bp^2}{2m} + U(\br) - \mu_2 \right ] \psi_2(\br) \; +  \;
w_2^2 \Phi_0 \psi_2^\dgr(\br) \psi_2(\br) \psi_2(\br) \; .
\ee
In these equations, again for brevity, the time dependence is not written
explicitly.

When considering a stationary state, it is possible to introduce an
effective temperature $T_{eff}$ characterizing the stationary input of
energy into the system [72-74]. Then an averaged behaviour of the latter,
in the sense of averaging over the observation time, can be described in
terms of effective statistical ensembles and the corresponding
thermodynamic quantities [75,76]. For stationary classical turbulence,
a similar approach has been used by Kraichnan and Montgomery [77].

For the considered case of the stationary quantum turbulence, we can
define the grand thermodynamic potential
\be
\label{37}
 \Om = - T_{eff}\ln \; {\rm Tr}\; \exp(-\bt\widetilde H) \;  ,
\ee
in which $\beta \equiv 1/T_{eff}$. The condition of stability for the
heterovortex mixture requires that the thermodynamic potential (37) be
minimal with respect to the corresponding geometric weights:
\be
\label{38}
 \frac{\dlt\Om}{\dlt w_1(\vom)} = 0 \; , \qquad
\frac{\dlt^2\Om}{\dlt w_1^2(\vom)} \; > \; 0  \; ,
\ee
where, to take into account normalization (27), one has to set
\be
\label{39}
w_2 = 1 - \int w_1(\vom) \; dm(\vom) \;  .
\ee

The first of conditions (38) gives
\be
\label{40}
\lgl \; \frac{\dlt\widetilde H}{\dlt w_1(\vom)} \; \rgl = 0 \;  ,
\ee
and the second yields
\be
\label{41}
\lgl \; \frac{\dlt^2\widetilde H}{\dlt w_1^2(\vom)} \; \rgl \; > \;
\bt \lgl \left ( \frac{\dlt\widetilde H}{\dlt w_1(\vom)}
\right )^2 \rgl \;  .
\ee
Since the right-hand side of inequality (41) is positive, the necessary
stability condition reads as
\be
\label{42}
\lgl\; \frac{\dlt^2\widetilde H}{\dlt w_1^2(\vom)} \; \rgl \; > \; 0  .
\ee

Let us introduce the notations
$$
K(\vom) \equiv \int \lgl \; \hat\psi(\vom,\br) \left [
\frac{\hat\bp^2}{2m} + U(\br) \right ]
\hat\psi(\vom,\br)\; \rgl \; d\br \; ,
$$
\be
\label{43}
\Phi(\vom) \equiv \Phi_0 \int
\lgl \; \hat\psi^\dgr(\vom,\br) \hat\psi^\dgr(\vom,\br)
\hat\psi(\vom,\br) \hat\psi(\vom,\br)\; \rgl \; d\br
\ee
for the superfluid state with vorticity $\vec{\omega}$ and
$$
K_2\equiv \int \lgl\; \psi_2^\dgr(\br) \left [
\frac{\hat\bp^2}{2m} + U(\br) \right ] \psi_2(\br)\; \rgl \; d\br \; ,
$$
\be
\label{44}
\Phi_2 \equiv \Phi_0 \int
\lgl \; \psi_2^\dgr(\br) \psi_2^\dgr(\br)
\psi_2(\br)\psi_2(\br)\; \rgl \; d\br
\ee
for the normal state. And let us define the integrals
\be
\label{45}
R_0(\vom) \equiv \int \rho_0(\vom,\br) \; d\br \; ,
\qquad R_1(\vom) \equiv \int \rho_1(\vom,\br) \; d\br \; ,
\qquad R_2 \equiv \int \rho_2(\br) \; d\br \; .
\ee
Then the numbers of condensed and uncondensed atoms in the $\vec{\omega}$
superfluid state, and atoms in the normal state are
\be
\label{46}
N_0(\vom) = w_1(\vom) R_0(\vom) \; , \qquad
N_1(\vom) = w_1(\vom) R_1(\vom) \;  , \qquad N_2 = w_2 R_2 \; ,
\ee
respectively. Also, we define
\be
\label{47}
\widetilde K_1(\vom) \equiv K(\vom) -\mu_0(\vom) R_0(\vom) -
\mu_1(\vom) R_1(\vom) \; , \qquad
\widetilde K_2 \equiv K_2 -\mu_2 R_2 \; .
\ee
Employing these notations, from Eq. (40), we find the geometric weight
for the superfluid state with vorticity $\vec{\omega}$ in the form
\be
\label{48}
w_1(\vom) =
\frac{\Phi_2+\widetilde K_2-\widetilde K_1(\vom)}{\Phi(\vom)+\Phi_2} \;  .
\ee
>From the stability condition (42) we obtain
\be
\label{49}
\Phi(\vom) + \Phi_2 > 0 \; .
\ee

The system chemical potential can be found from the relations
\be
\label{50}
F = \Om + \mu N = \Om + \int \left [\; \mu_0(\vom) N_0(\vom) +
\mu_1(\vom) N_1(\vom)\; \right ] dm(\vom) + \mu_2 N_2 \;  ,
\ee
defining the free energy. This yields
\be
\label{51}
\mu = \int \left [\; \mu_0(\vom) n_0(\vom) +
\mu_1(\vom) n_1(\vom) \; \right ] dm(\vom) + \mu_2 n_2 \;  ,
\ee
where the corresponding atomic fractions are
\be
\label{52}
n_0(\vom) \equiv \frac{N_0(\vom)}{N} \; \qquad
n_1(\vom) \equiv \frac{N_1(\vom)}{N} \; , \qquad
n_2 \equiv \frac{N_2}{N} \; .
\ee

The superfluid and normal phases are in mutual equilibrium, because
of which their chemical potentials are connected with each other. The
relation between the chemical potentials follows from the condition
of equilibrium $\delta F = 0$, where the variation is with respect
to the numbers of atoms. This is equivalent to the equation
\be
\label{53}
\int \left [\; \frac{\dlt F}{\dlt N_0(\vom)} \; \dlt N_0(\vom)
+ \frac{\dlt F}{\dlt N_1(\vom)} \; \dlt N_1(\vom)\; \right ]
dm(\vom) \; + \; \frac{\dlt F}{\dlt N_2} \; \dlt N_2 = 0 \;  .
\ee
Keeping the total number of atoms $N$ fixed implies that
\be
\label{54}
\int [ \dlt N_0(\vom) + \dlt N_1(\vom) ] \; dm(\vom) \; +
\; \dlt N_2 = 0 \;  .
\ee
By the meaning of the chemical potentials,
\be
\label{55}
\frac{\dlt F}{\dlt N_0(\vom)} = \mu_0(\vom) \; , \qquad
\frac{\dlt F}{\dlt N_1(\vom)} = \mu_1(\vom) \; , \qquad
\frac{\dlt F}{\dlt N_2} = \mu_2 \;  .
\ee
And for the atomic fractions, we have
\be
\label{56}
\frac{\dlt N_0(\vom)}{\dlt N} = n_0(\vom) \; , \qquad
\frac{\dlt N_1(\vom)}{\dlt N} = n_1(\vom) \; , \qquad
\frac{\dlt N_2}{\dlt N} = n_2 \; .
\ee
With the notation
\be
\label{57}
n_0 \equiv \int n_0(\vom) \; dm(\vom) \; , \qquad
n_1 \equiv \int n_1(\vom) \; dm(\vom) \;   ,
\ee
normalization (25) becomes
\be
\label{58}
 n_0 + n_1 + n_2 = 1\;  .
\ee

Invoking the above relations, from the equilibrium condition (53), we obtain
\be
\label{59}
 \mu_2 =
\frac{\int[\mu_0(\vom)n_0(\vom)+\mu_1(\vom)n_1(\vom)]dm(\vom)}{n_0+n_1}\; .
\ee
Substituting this into (51) gives
\be
\label{60}
\mu =
\frac{\int[\mu_0(\vom)n_0(\vom)+\mu_1(\vom)n_1(\vom)]dm(\vom)}{n_0+n_1}\; ,
\ee
that is,
\be
\label{61}
\mu=\mu_2 \;  .
\ee

The total superfluid density can be found [33] from the general expression
\be
\label{62}
\rho_s = \rho \; - \; \frac{\bt}{3mV} \; \Dlt^2(\hat{\bf P}) \;  ,
\ee
in which $\rho = N/V$ is the total atomic density and the momentum dispersion
$$
\Dlt^2(\hat\bP) \equiv \lgl \hat\bP^2 \rgl -
\lgl \hat\bP \rgl^2
$$
characterizes the amount of heat dissipated in the system. The total
momentum is defined as
\be
\label{63}
\hat\bP \equiv \lim_{u\ra 0} \; \frac{\prt\widetilde H_u}{\prt\bu}  \; ,
\ee
where $\widetilde{H}_u$ is the Hamiltonian for the system boosted with
velocity $u$ (see details in [33]). This gives
$$
\hat\bP = \hat\bP_1 + \hat\bP_2 \; , \qquad
\hat\bP_1 = \int \hat\bP(\vom) \; dm(\vom) \; ,
$$
$$
\hat\bP(\vom) = w_1(\vom)
\int \hat\psi(\vom,\br) (-i\nabla) \hat\psi(\vom,\br) \; d\br\; ,
$$
\be
\label{64}
\hat\bP_2 = w_2 \int \psi_2(\br) (-i\nabla) \psi_2(\br) \; d\br \; .
\ee
Because of the vanishing covariance
$$
{\rm cov} (\hat\bP_1,\hat\bP_2) \equiv
\frac{1}{2} \lgl \hat\bP_1 \hat\bP_2 + \hat\bP_2 \hat\bP_1 \rgl -
\lgl \hat\bP_1 \rgl \lgl \hat\bP_2 \rgl = 0 \; ,
$$
we have
$$
\Dlt^2(\hat\bP) = \Dlt^2(\hat\bP_1) + \Dlt^2(\hat\bP_2) \; .
$$
Therefore, the superfluid density (62) takes the form
\be
\label{65}
\rho_2 = \rho \; - \; \frac{\bt}{3mV} \left [ \Dlt^2(\hat\bP_1)
+ \Dlt^2(\hat\bP_2) \right ] \; .
\ee

Generally, a Bose system, subject to the action of an external alternating
field, can be in the following five states, which depend on the modulation
amplitude and time, that is, on the amount of energy pumped into the system.

(i) {\it Homogeneous superfluid}: If there is no external perturbation, the
system is in equilibrium, and the effective temperature equals the usual
temperature, $T_{eff} = T$. We assume that the latter is below the
Bose-Einstein condensation temperature, so that the system is Bose condensed
and is in a homogeneous superfluid state without vortices. The absence of
vortices implies zero vorticity $\vec{\omega} = 0$. The fact that the whole
system is filled by a homogeneous superfluid, without vortices, means that
the geometric weight
\be
\label{66}
w_1(0) = 1 \; .
\ee
This state remains under rather weak perturbations, when the effective
temperature is yet close to $T < T_c$.

(ii) {\it Vortex superfluid}: Disturbing the system by an alternating field
injects energy into the system. The pumped mechanical energy increases the
effective temperature. And, after sufficient amount of energy has been pumped
into the system, there appear quantized vortices. First, there arise just a
few of them, whose energy approximately equals the pumped energy. This can be
expressed through the sum
\be
\label{67}
 w_1(0) + \sum_i w_1(\vom_i) = 1 \;  ,
\ee
in which, in addition to the geometric weight $w_1(0)$, there appear a few
terms $w_1(\vec{\omega})$, with $\vec{\omega} \neq 0$, describing a small
(roughly speaking, less then ten) number of vortices.

(iii) {\it Turbulent superfluid}: When the energy, pumped into the system,
reaches a critical value, quantum turbulence develops with a great number of
quantized vortices forming a tangle. This can be denoted as
\be
\label{68}
\int w_1(\vom) \; dm(\vom) = 1 \; ,
\ee
the number of vortices being much larger than ten. The integral includes
the term $w_1(0)$.

(iv) {\it Mixture of turbulent superfluid and normal fluid}: A rather strong
pumping, not only produces vortices, but starts destroying superfluidity,
so that the admixture of the normal (nonsuperfluid) liquid appears. This
means that the equation
\be
\label{69}
\int w_1(\vom) \; dm(\vom) \; + \; w_2 = 1
\ee
becomes valid, with a nonzero geometric weight of the normal fluid $w_2 > 0$.

(v) {\it Normal fluid}: When the pumping becomes very strong, it can destroy
all Bose-Einstein condensate and, hence, superfluid, transferring the system
into the normal state. If there exists solely the normal phase, then
\be
\label{70}
 w_2 = 1 \; .
\ee
The normal state can be turbulent, but this would be the classical turbulence.

The overall picture is presented in the scheme of Fig. 1.

\section{Concluding remarks}

A statistical model is suggested, describing stationary states
arising in a superfluid subject to the action of an external
alternating field pumping energy into the system. The latter
is represented as a continuous mixture of dynamic phases with
different vorticities. The case of no vortices is included,
corresponding to zero vorticity. The possible destruction of
superfluidity by the pumping, leading to the developing normal
(nonsuperfluid) state is also taken into account. The vortices
arise owing to dynamic instability corresponding to the appearance
of unstable growing collective excitations.

The first three regimes, shown in the scheme of Fig. 1, that
is, homogeneous superfluid, vortex superfluid, and turbulent
superfluid, have been realized, by modulating the trapping
potential, in experiments [57,58] of the Bagnato group. A
detailed analysis of these experiments and their relation to
theory will be given in separate publications. Here, we give
some estimates for clarifying the physical picture by presenting
the typical parameters of a turbulent superfluid.

In experiments [57,58], trapped atoms of $^{87}$Rb are cooled
down to form Bose-Einstein condensate. Hence, atomic mass is
$m=1.445\times 10^{-22}$ g and scattering length is $a_s =0.577
\times 10^{-6}$ cm. Almost all atoms have been condensed,
composing the condensate of $N = 2 \times 10^5$ atoms, thermal
fraction being rather small. A cylindrical harmonic trap is
characterized by the radial frequency $\om_{\perp}=2\pi
\times 210$ Hz $ = 1.319\times 10^3$ s$^{-1}$ and longitudinal
frequency $\omega_z = 2 \pi \times 23$ Hz $ = 1.445 \times
10^2$ s$^{-1}$. The related oscillator lengths,
$$
l_\perp \equiv \sqrt{\frac{\hbar}{m\om_\perp} } \; ,
\qquad  l_z \equiv \sqrt{\frac{\hbar}{m\om_z} } \;  ,
$$
are $l_{\perp} = 0.744 \times 10^{-4}$ cm and $l_z = 2.248
\times 10^{-4}$ cm. The average frequency and oscillator length,
$$
\om_0 \equiv \left ( \om_\perp^2 \om_z \right )^{1/3} \; ,
\qquad l_0 \equiv \left ( l_\perp^2 l_z \right )^{1/3} =
\sqrt{\frac{\hbar}{m\om_0} }\; ,
$$
are $\om_0=0.631\times 10^3$ s$^{-1}$ and $l_0=1.076\times 10^{-4}$ cm.
The effective volume of the condensate
$$
V_{eff} \equiv \pi l_\perp^2 2 l_z = 2\pi l_0^3
$$
is $V_{eff}=0.783\times 10^{-11}$ cm$^3$. Therefore, the average
condensate density is $\rho\equiv N/V_{eff}=2.554\times 10^{15}$
cm$^{-3}$. The average interatomic distance is $a=\rho^{-1/3}=0.732
\times 10^{-5}$ cm. This shows that $a_s\ll a$ and the gas parameters
are small: $\rho a_s^3=0.491\times 10^{-3}$ and $\rho^{1/3} a_s=0.079$.
That is, the atomic interactions are weak.

The trap is subject to an external field modulation during the
time $t_{ext}=0.02$ s $- 0.06$ s with an alternating potential of
frequency $\omega_{mod} =2 \pi\times 200$ Hz $ = 1.257\times 10^3$ s$^{-1}$.
Thence the modulation period is $t_{mod}\equiv 2\pi/\omega_{mod}=5
\times 10^{-3}$ s. The local-equilibrium time [29] $t_{loc}=m/\hbar
\rho a_s=0.929\times 10^{-4}$ s is much shorter than the modulation
period $t_{mod}$, which implies that the system is always in local
equilibrium.

The typical size of a vortex core is the healing length
$$
 \xi = \frac{1}{\sqrt{2m\rho\Phi_0} } =
\frac{1}{\sqrt{8\pi\rho a_s} } \;  ,
$$
which is $\xi=0.519\times 10^{-5}$ cm, being of order of interatomic
distance $a$.

The sound velocity
$$
c = \sqrt{\frac{\rho}{m}\;\Phi_0 } = \frac{\hbar}{m} \;
\sqrt{4\pi\rho a_s}
$$
becomes $c=0.994$ cm$/$s. This defines the coherence length $l_{coh}=
\hbar/m c = 0.735 \times 10^{-5}$ cm that is of order of the healing
length.

The vortex velocity (1), for $n=1$, is of order
$$
v= \frac{\hbar}{2ml_0}\; \ln\left ( \frac{l_0}{\xi} \right ) \; ,
$$
which gives $v = 0.103$ cm$/$s.

Another typical velocity is that characterizing the velocity of atomic
collisions $v_{col} = \hbar/m a_s = 12.65$ cm$/$s during the collision
time $t_{col} = a_s/v_{col} = 0.456 \times 10^{-7}$ s. Thus, the
relation between the characteristic velocities is
$$
v \ll c \ll v_{col} \;  .
$$

The characteristic vortex energy
$$
\ep_{vor} = \frac{\hbar}{ml_0} \;
\ln \left ( \frac{l_0}{\xi} \right )
$$
is $\varepsilon  = 1.912 \times 10^3$ s$^{-1}$.

Summarizing the relations between the typical times yields
$$
 t_{col} \ll t_{loc} \ll t_{mod} \ll t_{ext} \; .
$$
And the relations between the characteristic lengths are
$$
a_s \ll \xi \sim a \sim l_{coh} \ll l_0 \;  .
$$

Finally, the turbulent regime has to be formed by many vortices,
$N_{vor} > 10$. But there is the maximal number of vortices of order
$$
N_{max} \sim \left ( \frac{l_0}{l_{coh} }\right )^2 \; \div \;
\left ( \frac{l_0}{\xi} \right )^2 \;  ,
$$
before the vortices start essentially overlapping with each other.
If this overlap becomes too strong, the superfluid state is destroyed.
Therefore the number of vortices in a turbulent superfluid is in the
interval
$$
 10 < N_{vor} < 200 \; \div \; 400 \; .
$$
This is in agreement with experiment [58], where up to 200 vortices
were observed.

The turbulent superfluid, being a tangled mixture of many condensates
with vortices, should exhibit a specific behaviour, when the trapping
potential is switched off. Each particular condensate, being released 
from the trap, expands anisotropically. But, since the mixture of many 
condensates, as a whole, is isotropic, all particular anisotropic 
directions are averaged out, so that the total turbulent cloud expands 
isotropically, keeping its aspect ratio during the whole free expansion.
This effect was observed in experiment [58].

As is mentioned above, a more detailed analysis of these and related
experiments will be given in separate publications.

\vskip 5mm

{\it Acknowledgements}. I am grateful to V.S. Bagnato for discussions
of experiments [57,58]. Financial support of the Russian Foundation
for Basic Research is appreciated.

\newpage

\section*{Appendix. Continuous products}

The model of quantum turbulence in superfluids, suggested in this
paper, is based on the notion of continuous products. For the
self-consistency of the paper, in this Appendix, the main definitions
are given, related to continuous products. Such products have been
used for treating continuous heterophase mixtures [63,64] and models
of continuous random walk [78,79].

Let a complex function $f(x): \mathbb{R} \rightarrow \mathbb{C}$ of
a real variable $x \in \mathbb{R}$ be given. Suppose, we consider
$x$ in an interval $[a,b]$. Divide this interval on $n$ parts of
length $\Dlt x_i$ so that $\sum_{i=1}^n \Delta x_i = b-a$. Define
the limiting procedure
$$
n \ra \infty \; , \qquad \Dlt x_i \ra 0 \qquad
(i=1,2,\ldots,n) \;  .
$$

A {\it definite continuous product} of the function $f(x)$ on an
interval $[a,b]$ is given by the limit
$$
\prod_{x=a}^b f(x) \equiv
\lim_{n\ra\infty} \prod_{i=1}^n f(a+\Dlt x_i) \; , \hspace{2cm}  (A.1)
$$
where the above limiting procedure is assumed.

In order to transform this limit to a more tractable form, let us take
the logarithm of the limiting equation, which gives
$$
\log  \prod_{x=a}^b f(x) = \lim_{n\ra\infty} \sum_{i=1}^n
\log f(a+\Dlt x_i) \; .
$$
The logarithm can be taken over any base. Notice that the right-hand
side here is the definition of an integral, so that
$$
\log  \prod_{x=a}^b f(x) = \int_a^b \log f(x) \; dm(x) \;  ,
$$
where $m(x)$ is a measure on $\mathbb{R}$. Since the logarithm can
be taken over any base, we may chose the natural logarithm. Then
exponentiation yields another form of the definition for a
{\it definite continuous product}:
$$
\prod_{x=a}^b f(x) \equiv \exp\left \{
\int_a^b \ln f(x) \; dm(x) \right \} \;  .  \hspace{2cm} (A.2)
$$
The logarithm of a complex function is given by the expression
$$
\ln f(x) = \ln | f(x)| + i\; {\rm arg} f(x) \;  .
$$

Using the notion of indefinite integrals, it is possible to define an
{\it indefinite continuous product}
$$
\prod_x f(x) = c \exp \left \{ \int \ln f(x) \; dm(x)
\right \} \;  ,  \hspace{2cm} (A.3)
$$
where $c > 0$ is a positive constant. Then, we can introduce the action
inverse to the continuous product as
$$
f(x) = \exp \left \{ \frac{1}{m'(x)} \; \frac{d}{dx} \;
\ln \prod_x f(x) \right \} \;  ,
$$
where $m'(x) \equiv dm(x)/dx$.

It is straightforward to generalize the notion of a continuous
product as follows. Let a complex function $f(x): \mathbb{D}
\rightarrow \mathbb{C}$ be given on a measurable domain
$\mathbb{D}$ with a measure $m(x)$. Then the {\it continuous product
of $f(x)$ over the domain $\mathbb{D}$} is
$$
 \prod_{x\in\mathbb{D} } f(x) \equiv \exp \left \{
\int_{x\in\mathbb{D} } \ln f(x) \; dm(x) \right \} \; .
\hspace{2cm} (A.4)
$$

It is possible to introduce continuous products of spaces
[63,64,80]. Let $y_x, z_x \in {\cal U}_x$ be vectors of a unitary
space ${\cal U}_x$ with a scalar product $(y_x,z_x)$ and
$x\in\mathbb{D}$. The manifold $\mathbb{D}$ is assumed to be
measurable, with a measure $m(x)$. The norm of $y_x$, generated
by the scalar product, is $\|y_x\| \equiv\sqrt{(y_x,y_x)}$.

The {\it continuous tensor product of unitary spaces},
$$
{\cal U} \; \equiv \; \bigotimes_{x\in\mathbb{D} } {\cal U}_x
\hspace{7cm}     (A.5)
$$
consists of the vectors that are continuous tensor products
$$
y \; \equiv \; \bigotimes_{x\in\mathbb{D} } y_x \; \in \; {\cal U} \;  .
\hspace{6cm}  (A.6)
$$
The scalar product in $\cal{U}$ is given by
$$
(y,z) \equiv \prod_{x\in\mathbb{D} } (y_x,z_x) =
\exp\left\{ \int_\mathbb{D} \ln(y_x,z_x) \; dm(x)
\right \} \;  . \hspace{1cm}  (A.7)
$$
Respectively, the norm of $y\in{\cal U}$, generated by the scalar
product, is
$$
|| y || \equiv \sqrt{ (y,y) } = \exp\left\{
\int_\mathbb{D} \ln || y_x || \; dm(x) \right \} \; .
$$
When ${\cal U}_x$ is complete, it becomes a Hilbert space (complete
unitary space). Then $\cal{U}$ is also a Hilbert space.

Finally, one can introduce continuous tensor products of operators.
Let ${\cal H}_x$ be a Hilbert space, with $x \in \mathbb{D}$, where
$\mathbb{D}$ is measurable with a measure $m(x)$. And let an operator
$\hat{A}_x$ be given on ${\cal H}_x$. The operator norm can be defined
as
$$
|| \hat A_x || \equiv \sup_{y_x \in {\cal H}_x} \;
\frac{|| \hat A_x y_x ||}{|| y_x || } \qquad
(y_x \neq 0 ) \;  .
$$

The continuous tensor product of the Hilbert spaces ${\cal H}_x$ is
$$
{\cal H} \equiv \bigotimes_{x\in\mathbb{D}} {\cal H}_x \;   ,
$$
similarly to (A.5). An operator $\hat{A}$ on $\cal{H}$ is an
{\it operator continuous tensor product}
$$
\hat A \equiv  \bigotimes_{x\in\mathbb{D}}
\hat A_x \; ,  \hspace{4cm} (A.8)
$$
whose action on $y \in \cal{H}$ is given by the tensor product
$$
 \hat A_y y \equiv \bigotimes_{x\in\mathbb{D}}
\hat A_x y_x\; .  \hspace{4cm} (A.9)
$$
The operator norm of $\hat{A} \in \cal{H}$, defined in the standard
way,
$$
|| \hat A || \equiv \sup_{y\in{\cal H} } \;
\frac{||\hat A y ||}{||y ||} \qquad (y \neq 0) \;  ,
$$
results in the relation
$$
 || \hat A || = \exp \left \{ \int_\mathbb{D}
\ln || \hat A_x || \; dm(x) \right \} \; . \hspace{1cm}  (A.10)
$$

These definitions and properties of continuous products allow us to
accomplish all necessary calculations for the considered turbulent superfluid
represented as a continuous vortex mixture, in which the role of the index $x$
is played by the vorticity $\vec{\omega}$.

\newpage

\begin{figure}[h]
\centerline{\includegraphics[width=12cm]{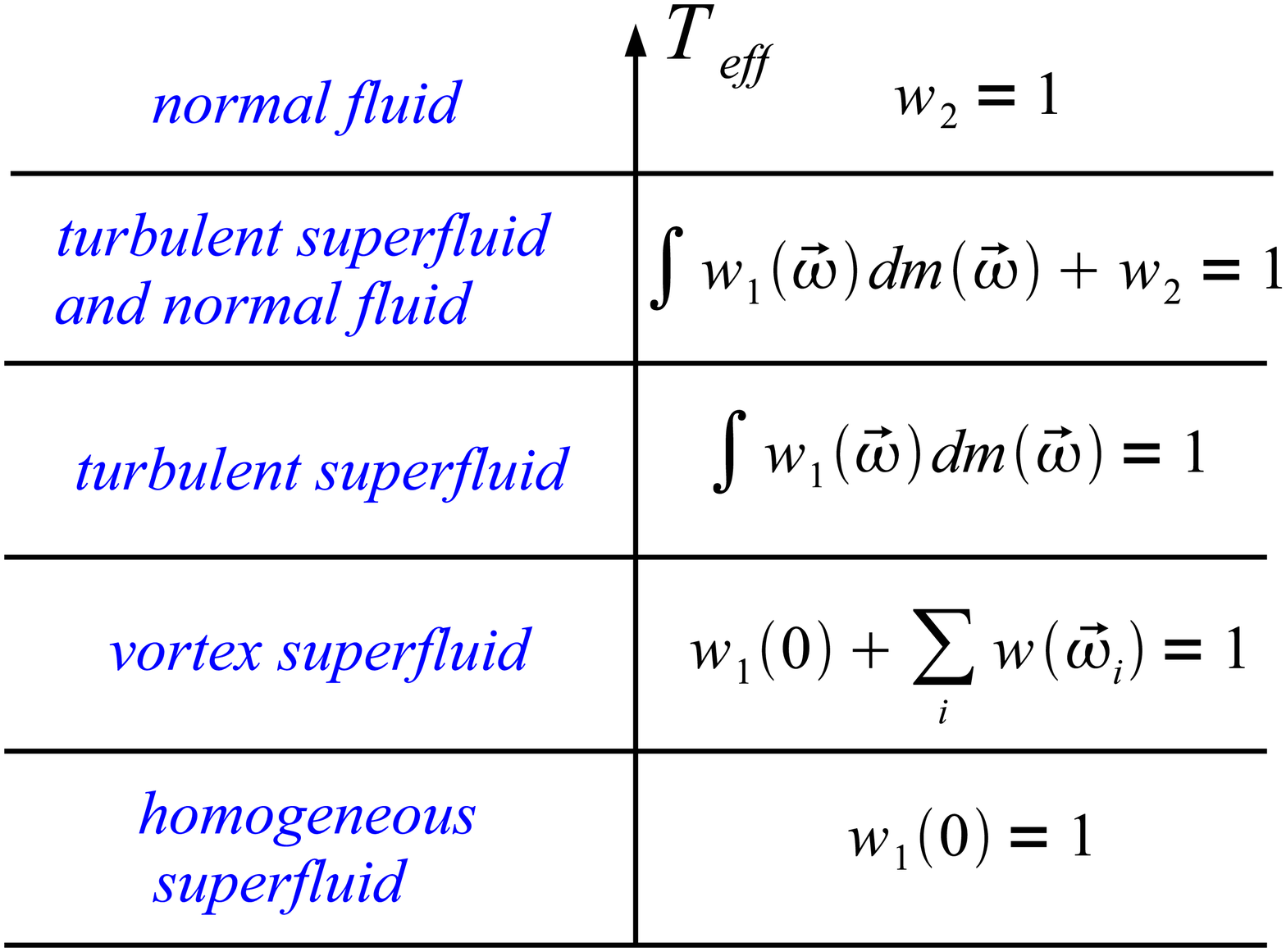}}
\caption{Scheme of the sequence of states for a superfluid
subject to the action of an alternating external field.}
\label{fig:Fig.1}
\end{figure}

\end{document}